\documentclass[12pt]{article}

\makeatletter

\newif\ifCrossDocumentPackageAvailable
\IfFileExists{xr-hyper.sty}{%
  \usepackage{xr-hyper}%
  \CrossDocumentPackageAvailabletrue
}{%
  \IfFileExists{xr.sty}{%
    \usepackage{xr}%
    \CrossDocumentPackageAvailabletrue
  }{%
    \CrossDocumentPackageAvailablefalse
  }%
}

\newif\ifStaticCrossDocumentReferences
\StaticCrossDocumentReferencesfalse

\newcommand{\CrossDocumentStaticLabel}[4]{%
  \expandafter\protected@xdef\csname r@#1\endcsname
    {{#2}{#3}{}{#4}{\CrossDocumentPDF}}%
}

\newcommand{\CrossDocumentUseFallback}[3]{%
  \PackageWarningNoLine{cross-document-refs}{%
    Using static references for `#1' because #2}%
  \def\CrossDocumentPDF{#1.pdf}%
  #3%
}

\newcommand{\LoadCrossDocumentReferences}[2]{%
  \ifStaticCrossDocumentReferences
    \CrossDocumentUseFallback{#1}{the static-reference mode was requested}{#2}%
  \else
    \IfFileExists{#1.aux}{%
      \ifCrossDocumentPackageAvailable
        \begingroup
          \let\bibcite\@gobbletwo
          \externaldocument{#1}%
        \endgroup
      \else
        \CrossDocumentUseFallback{#1}{neither xr-hyper nor xr is available}{#2}%
      \fi
    }{%
      \CrossDocumentUseFallback{#1}{`#1.aux' was not found}{#2}%
    }%
  \fi
}

\newcommand{\LoadSupplementReferences}{%
  \LoadCrossDocumentReferences{supp-fpca-online}{%
    \CrossDocumentStaticLabel{eq:taukbeta}{S5}{21}{equation.5.5}%
    \CrossDocumentStaticLabel{fig:aqs-sites-binned}{S11}{64}{figure.caption.29}%
    \CrossDocumentStaticLabel{fig:fpca1d-taupath}{S9}{63}{figure.caption.28}%
    \CrossDocumentStaticLabel{fig:gfr-eigfun}{S12}{65}{figure.caption.30}%
    \CrossDocumentStaticLabel{lem:difflogtau}{S1}{20}{lemma.1}%
    \CrossDocumentStaticLabel{rmk:altn-explore}{S1}{21}{remark.1}%
    \CrossDocumentStaticLabel{sup:ass}{S5}{19}{section.5}%
    \CrossDocumentStaticLabel{sup:bspline}{S2}{10}{section.2}%
    \CrossDocumentStaticLabel{sup:ci-online}{S7.5}{39}{subsection.7.5}%
    \CrossDocumentStaticLabel{sup:conv}{S6}{24}{section.6}%
    \CrossDocumentStaticLabel{sup:expr-details}{S9.1}{42}{subsection.9.1}%
    \CrossDocumentStaticLabel{sup:extra-figs}{S9.9}{58}{subsection.9.9}%
    \CrossDocumentStaticLabel{sup:hess}{S3.2}{13}{subsection.3.2}%
    \CrossDocumentStaticLabel{sup:manifold}{S1}{4}{section.1}%
    \CrossDocumentStaticLabel{sup:mean}{S8}{40}{section.8}%
    \CrossDocumentStaticLabel{sup:set-cwb}{S9.3.1}{46}{subsubsection.9.3.1}%
  }%
}

\makeatother

\LoadSupplementReferences

\newcommand{\blind}{0}

\usepackage[utf8]{inputenc}
\usepackage{hyperref}
\hypersetup{
colorlinks=true,
linkcolor=blue,
filecolor=blue,
citecolor = blue,
urlcolor=cyan,
}
\usepackage[dvipsnames]{xcolor}

\usepackage{url} 
\usepackage{graphicx}
\graphicspath{ {figures/} }
\usepackage{caption}
\usepackage{subcaption}
\usepackage{tabularx}
\usepackage[flushleft]{threeparttable}
\usepackage{multirow}
\usepackage{booktabs}
\usepackage[detect-all]{siunitx}
\usepackage{etoolbox} 
\newrobustcmd{\B}{\bfseries} 
\usepackage{makecell}
\usepackage[normalem]{ulem}
\usepackage[ruled]{algorithm2e}
\usepackage{algpseudocode}
\SetKwInput{KwInit}{Initialize}
\SetKwInput{KwInput}{Input}
\SetKwInput{KwOutput}{Output}
\SetKwComment{Comment}{/* }{ */}

\usepackage{amsmath}
\usepackage{amsthm,amssymb,bm,bbm}
\usepackage{mathtools} 
\allowdisplaybreaks[2]

\usepackage{enumitem}
\usepackage{authblk}

\usepackage{natbib}

\newcommand{\E}{\mathbb{E}}

\newcommand{\cov}{\text{cov}}

\newcommand{\bb}{\mathbf{b}}

\newcommand{\bt}{\mathbf{t}}
\newcommand{\bs}{\mathbf{s}}

\newcommand{\bv}{\mathbf{v}}

\newcommand{\by}{\mathbf{y}}

\newcommand{\bB}{\mathbf{B}}
\newcommand{\bC}{\mathbf{C}}

\newcommand{\bG}{\mathbf{G}}

\newcommand{\bI}{\mathbf{I}}

\newcommand{\bS}{\mathbf{S}}

\newcommand{\bU}{\mathbf{U}}
\newcommand{\bV}{\mathbf{V}}

\newcommand{\btheta}{{\bm{\theta}}}
\newcommand{\blambda}{{\bm{\lambda}}}

\newcommand{\bmeta}{{\bm{\eta}}}

\newcommand{\bDelta}{\mathbf{\Delta}}

\newcommand{\bTheta}{\mathbf{\Theta}}
\newcommand{\bSigma}{\mathbf{\Sigma}}

\newcommand{\bPsi}{\mathbf{\Psi}}

\newcommand{\calH}{\mathcal{H}}

\newcommand{\calN}{\mathcal{N}}
\newcommand{\calM}{\mathcal{M}}
\newcommand{\calP}{\mathcal{P}}
\newcommand{\calR}{\mathcal{R}}
\newcommand{\calS}{\mathcal{S}}
\newcommand{\calT}{\mathcal{T}}

\newcommand{\rmd}{\mathrm{d}}
\newcommand{\rmtr}{\mathrm{tr}}
\newcommand{\rmDiag}{\mathrm{Diag}}

\newcommand{\rmSt}{\mathrm{St}}

\DeclareMathSymbol{\m}{\mathbin}{AMSa}{"39}

\newcommand{\pp}{\partial}
\newcommand{\stg}{\rmSt_{\bG}(R, p)}

\theoremstyle{plain}
\newtheorem{assumption}{Assumption}
\newtheorem{theorem}{Theorem}

\makeatletter
\let\save@mathaccent\mathaccent
\newcommand*\if@single[3]{%
 \setbox0\hbox{${\mathaccent"0362{#1}}^H$}%
 \setbox2\hbox{${\mathaccent"0362{\kern0pt#1}}^H$}%
 \ifdim\ht0=\ht2 #3\else #2\fi
 }
\newcommand*\rel@kern[1]{\kern#1\dimexpr\macc@kerna}
\newcommand*\widebar[1]{\@ifnextchar^{{\wide@bar{#1}{0}}}{\wide@bar{#1}{1}}}
\newcommand*\wide@bar[2]{\if@single{#1}{\wide@bar@{#1}{#2}{1}}{\wide@bar@{#1}{#2}{2}}}
\newcommand*\wide@bar@[3]{%
 \begingroup
 \def\mathaccent##1##2{%
 \let\mathaccent\save@mathaccent
 \if#32 \let\macc@nucleus\first@char \fi
 \setbox\z@\hbox{$\macc@style{\macc@nucleus}_{}$}%
 \setbox\tw@\hbox{$\macc@style{\macc@nucleus}{}_{}$}%
 \dimen@\wd\tw@
 \advance\dimen@-\wd\z@
 \divide\dimen@ 3
 \@tempdima\wd\tw@
 \advance\@tempdima-\scriptspace
 \divide\@tempdima 10
 \advance\dimen@-\@tempdima
 \ifdim\dimen@>\z@ \dimen@0pt\fi
 \rel@kern{0.6}\kern-\dimen@
 \if#31
 \overline{\rel@kern{-0.6}\kern\dimen@\macc@nucleus\rel@kern{0.4}\kern\dimen@}%
 \advance\dimen@0.4\dimexpr\macc@kerna
 \let\final@kern#2%
 \ifdim\dimen@<\z@ \let\final@kern1\fi
 \if\final@kern1 \kern-\dimen@\fi
 \else
 \overline{\rel@kern{-0.6}\kern\dimen@#1}%
 \fi
 }%
 \macc@depth\@ne
 \let\math@bgroup\@empty \let\math@egroup\macc@set@skewchar
 \mathsurround\z@ \frozen@everymath{\mathgroup\macc@group\relax}%
 \macc@set@skewchar\relax
 \let\mathaccentV\macc@nested@a
 \if#31
 \macc@nested@a\relax111{#1}%
 \else
 \def\gobble@till@marker##1\endmarker{}%
 \futurelet\first@char\gobble@till@marker#1\endmarker
 \ifcat\noexpand\first@char A\else
 \def\first@char{}%
 \fi
 \macc@nested@a\relax111{\first@char}%
 \fi
 \endgroup
}
\makeatother

\newcommand{\wb}{\widebar}

\begin{document}

\def\spacingset#1{\renewcommand{\baselinestretch}%
	{#1}\small\normalsize} \spacingset{1}


\if0\blind
{
\title{\bf Online Learning of Functional Principal Component Analysis for Multidimensional Functional Data}
\author[1,2]{Muye Nanshan}
\author[1,3]{Nan Zhang\thanks{NZ was supported by Grant No. LGL-1987 of the LinGang Laboratory.}}
\author[2]{Jiguo Cao\thanks{JC was supported by a discovery grant (RGPIN-2023-04057) of the Natural Sciences and Engineering Research Council of Canada (NSERC) and the Canada Research Chairs program.}}
\affil[1]{School of Data Science, Fudan University, Shanghai, China}
\affil[2]{Department of Statistics and Actuarial Science, Simon Fraser University, Burnaby, Canada}
\affil[3]{Institute of Science and Technology for Brain-Inspired Intelligence, Fudan University, Shanghai, China}
\date{}
	\maketitle
} \fi
\if1\blind
{
	\bigskip
	\bigskip
	\bigskip
	\begin{center}
		{\LARGE\bf Online Learning of Functional Principal Component Analysis for Multidimensional Functional Data}
	\end{center}
	\medskip
} \fi

\bigskip
\begin{abstract}
Multidimensional functional data streams arise in diverse scientific fields, yet their analysis poses significant challenges. We propose a novel online framework for functional principal component analysis that enables efficient and scalable modeling of such data. Our method represents functional principal components using tensor product splines, enforcing smoothness and orthonormality through a penalized framework on a Stiefel manifold. We develop an efficient Riemannian stochastic gradient descent algorithm and a Riemannian adaptive gradient (AdaGrad) variant, both utilizing iterative averaging techniques to stabilize the estimation and accelerate convergence. Additionally, a dynamic tuning strategy for smoothing parameter selection is developed based on a rolling averaged block validation score that adapts to the streaming nature of the data. Furthermore, we derive the asymptotic normality of the estimators and construct pointwise confidence intervals to quantify the uncertainty of the estimated functional principal components. Extensive simulations and real-world applications demonstrate the flexibility and effectiveness of this framework for analyzing multidimensional functional data.
\end{abstract}

\noindent%
\vfill

\newpage
\spacingset{1.8} 

\section{Introduction}\label{sec:intro}

Functional principal component analysis (FPCA) is a widely used technique in functional data analysis \citep{ramsay2005functional,hsing2015theoretical}. As an extension of traditional PCA for multivariate data, FPCA aims to identify a lower-dimensional subspace capturing the major variability in a sample of realizations from a random process. Infinite-dimensional functional objects are represented by a few leading functional principal components, not only providing better visualization and interpretability but also facilitating subsequent statistical analyses such as clustering, classification, and regression.

In the big data era, modern advances in data acquisition bring new challenges in analyzing functional data of complex forms. The primary focus of this work is the online learning of functional principal components for multidimensional functional data that arrive in a streaming fashion.
One of our motivating examples is the daily Air Quality Index (AQI) collected by the United States Environmental Protection Agency, where spatial maps are recorded over 3,000 monitoring sites over 40 years. While classical methods for functional data analysis are well-suited for simple curves, they encounter a severe computational bottleneck when applied to multidimensional random fields. Specifically, traditional {\it batch learning} procedures require the construction and decomposition of a high-dimensional covariance tensor to represent the underlying process. The memory used to store and estimate this batch covariance structure grows prohibitively. In such cases, online learning becomes a practical necessity. By sequentially processing data and incrementally updating model parameters, online algorithms provide a feasible alternative to large-scale multidimensional functional data analysis by bypassing the daunting computational overhead.



Beyond the constraints of computational scale, the benefit of developing an online framework is further motivated by the growing importance of data security and privacy in medical longitudinal research. Our second motivating application involves Glomerular Filtration Rate (GFR) trajectories, which tracks kidney function for over 140,000 transplant recipients \citep{liu2023functional}. In such massive cohorts, new data are frequently added as more patients are monitored over time. Retraining a batch model from scratch when new records are added is not only time-consuming but also introduces significant data security concerns. The potential data breach is recently highlighted as major challenges in the international biobank repositories. In contrast, online learning provides a practical solution by allowing model estimates to be incrementally updated. By processing newly arrived data in manageable data blocks, the proposed framework avoids repeated access to the entire database for every update. Its sequential nature ensures that the analysis remains up-to-date while protecting data privacy by reducing the amount of raw and private information held in the active computing environment at any given time.



In the literature on batch learning of functional principal components (FPCs), there are two typical approaches. The first follows a two-step procedure in which FPCs are extracted via eigenanalysis after a smooth estimate of the covariance function is obtained. Although there have been many works on covariance function estimation for one-dimensional functional data \citep{yao2005functional, xiao2018fast} and a few recent works on multidimensional functional data \citep{zhou2014principal,wang2022low,sarkar2022covnet}, such two-stage procedures suffer from the estimation quality of covariance functions. In particular, when the input domain is of dimension $d$, it is difficult to estimate $2d$-dimensional covariance surfaces due to the curse of dimensionality. The second approach takes a more direct strategy. It approximates the FPCs with basis functions and models the functional observations by linear combinations of the FPCs. Parameters such as basis coefficients are estimated under the likelihood framework with possible model complexity penalization. A partial list of such work includes \cite{james2000principal}, \cite{yao2005functional}, \cite{peng2009geometric}, \cite{lila2016smooth} and \cite{he2024penalized}.
In addition to avoiding the daunting covariance function estimation, this approach enjoys explicit basis representations for the FPCs, and it is convenient to introduce regularities such as smoothness or orthogonality.

Our contribution in this article is threefold.
First, we propose an online learning method to estimate the leading Functional Principal Components (FPCs) for sequentially arriving multidimensional functional data. We approximate the leading FPCs within a low-dimensional space of smooth functions and directly represent the FPCs with tensor product splines, thereby avoiding the challenges associated with estimating the high-dimensional covariance function. To maintain the orthonormal properties of the eigenfunctions, the basis coefficients representing the FPCs are constrained to a Stiefel manifold. 
Our online FPCA method is developed within a penalized framework that combines smoothness regularization with the intrinsic geometric structure of the FPCs. Unlike existing batch FPCA methods that employ spline estimation without penalty \citep{james2000principal, peng2009geometric}, our approach explicitly controls model complexity in a continuous manner. It has been observed that, in nonparametric function estimation, penalized splines offer better performance compared to unpenalized polynomial regression \citep{huang2021asymptotic}.

Second, we design efficient Riemannian stochastic gradient descent (RSGD) algorithms to update model parameters with sequentially arriving data. Stochastic gradient descent (SGD) is a standard technique in machine learning. It naturally fits the online setting by providing a stochastic approximation to the gradients using the newly received data block \citep{bottou2018optimization}. Under our online FPCA formulation, we perform model parameter optimization along a generalized Stiefel manifold \citep{absil2008optimization,sato2019cholesky} to respect the intrinsic geometric structure of the parameter space. To further enhance practical performance, we develop a Riemannian adaptive gradient (RAdaGrad) method tailored for online FPCA by extending the work by \cite{duchi2011adaptive}. RAdaGrad updates gradients by scaling them using estimates of the second moments, providing generally more efficient updating directions compared to the RSGD. Both the RSGD and RAdaGrad algorithms utilize iterative averaging techniques \citep{polyak1992acceleration,tripuraneni2018averaging} to stabilize the estimation process and facilitate optimal convergence rates. 

%
Third, we introduce a dynamic tuning scheme for selecting an appropriate smoothing parameter throughout the online learning process. In nonparametric estimation of FPCs, tuning the smoothing parameter is crucial for balancing model bias and variance. While many online nonparametric methods keep regularization parameters fixed, the importance of dynamic tuning has been increasingly recognized. For instance, \cite{zhang2022sieve} explored nonparametric regression using a sieve-SGD estimator, allowing the sieve size to grow with the accumulated sample size. \cite{yang2021online} proposed an online local polynomial estimation of mean and covariance functions, employing a dynamic sequence of candidate bandwidths. To address this, we integrate the validation principle with the streaming nature of online learning. Specifically, the newly arrived data block serves as a validation set on a rolling basis. We define an averaged block validation score to select the optimal hyperparameter at the current stage and form a compact set of future candidates.
This dynamic tuning strategy maintains a small set of candidate smoothing parameters and updates them as more data blocks are processed.
Furthermore, we establish the asymptotic properties of the proposed framework.
We prove the convergence of the RSGD algorithm under dynamic tuning and derive the asymptotic properties of the iteratively averaged estimator. These theoretical results allow us to construct pointwise confidence intervals, offering uncertainty quantification for the estimated functional principal components.

The rest of this article is organized as follows.
Section~\ref{sec:omfpca} formulates the online FPCA framework and our penalized estimation framework. Section~\ref{sec:rsgd-fix} presents the RSGD algorithm with tuning parameter fixed. Section~\ref{sec:dyn-tuning} describes our dynamic tuning procedure to adapt the smoothness of the FPCs to sequentially arriving data blocks.
We establish the asymptotic properties of the RSGD algorithm in Section~\ref{sec:conv}. To further improve the practical performance, we develop in Section~\ref{sec:adagrad} the Riemannian adaptive gradient method. The numerical performance of the proposed method is evaluated by simulation studies in Section~\ref{sec:sim} and two real data applications in Section~\ref{sec:data}. Technical details are collected in the supplementary materials.

\section{Online Learning for Multidimensional FPCA}\label{sec:omfpca}

In this section, we present a new online learning framework for multidimensional FPCA. 
It employs regularized spline approximation for FPC estimation and formulates multidimensional FPCA as a manifold optimization problem under the online setting.

\subsection{FPC and Basis Representation}\label{sec:formulation}

Suppose that we sequentially obtain observations $\{\by_1, \dots, \by_N\}$, where $\by_i\in\mathbb R^{m_i}$ represents $m_i$ measurements of a function $X_i(\bt)$ defined on a $d$-dimensional domain $\calT$. The total sample size $N$ is allowed to be infinite, and the number of measurements for each subject, $m_i$, is uniformly bounded. The $j$th component of $\by_i$, denoted by $y_{ij}$, is the measurement of a function $X_i(\bt)$ on a random location $\bt_{ij}$ contaminated by a noise $\varepsilon_{ij}$, i.e., $y_{ij} = X_i(\bt_{ij}) + \varepsilon_{ij}$ for $1\leq j\leq m_i$, and the noises are assumed to be zero-mean and have a constant variance $\sigma^2$.

We assume that the underlying functions $X_1(\bt), \dots, X_N(\bt)$ are independent copies of a random field $X(\bt)\in L^2(\calT)$. The Mercer's theorem implies that the covariance function $\gamma(\bs,\bt)=\cov(X(\bs), X(\bt))$ admits a spectral decomposition $\sum_{r=1}^\infty \lambda_{r} \phi_{r}(\bs) \phi_{r}(\bt)$, where $\lambda_1\geq\lambda_2\geq\dots>0$ are the eigenvalues, and $\{\phi_{r}: r\geq 1\}$ are the corresponding orthonormal eigenfunctions in $L^2(\mathcal{T})$. The eigenfunctions capturing the major variation of $X(\bt)$ are referred to as functional principal components (FPCs) in the context of FPCA. Given a pre-specified rank $R$, the truncated expansion using the leading FPCs provides a low rank approximation to the covariance function such that
\begin{equation}\label{eq:cov-rankr}
	\gamma(\bs,\bt) \approx \sum_{r=1}^R \lambda_{r} \phi_{r}(\bs) \phi_{r}(\bt).
\end{equation}
Our goal is to learn the top $R$ FPCs with the sequentially arrived observations.

We represent the FPCs by spline basis approximations. When $\calT$ is a one-dimensional domain, B-spline basis is a popular choice due to its computational efficiency. It can be easily generalized to multidimensional case as the tensor product B-spline basis \citep{wood2017generalized}. See Section~\ref{sup:bspline} in the supplementary materials for the construction of the tensor product B-spline basis on a multidimensional domain.
Let $\bb(\bt)$ be the spline basis on $\calT$, consisting of $p$ basis functions. The $r$th FPC, $\phi_{r}$, is approximated by the basis expansion $\btheta_r^\top\bb(\bt)$ where $\btheta_r\in\mathbb R^p$ is the vector of basis coefficients. Let $\bTheta=(\btheta_1,\ldots,\btheta_R)\in \mathbb{R}^{p\times R}$ collect the basis coefficients of the top $R$ FPCs.
The estimation of FPCs is thereby translated into the estimation of spline coefficients $\bTheta$. Moreover, the orthonormal properties for the eigenfunctions, namely $\int \phi_r(\bt) \phi_s(\bt) \rmd \bt = \delta_{rs}$, where $\delta_{rs}$ is the Kronecker delta function, can be written as the constraint $\bTheta^\top \bG \bTheta = \bI_R$ with $\bG=\int \bb(\bt)\bb(\bt)^\top \rmd \bt$ being the Gram matrix. This constraint corresponds to a geometric structure referred to as a generalized Stiefel manifold \citep{peng2009geometric, sato2019cholesky},
$\stg = \{\bTheta\in\mathbb{R}^{p\times R}: \bTheta^\top \bG \bTheta = \bI_R\}, $
see Section~\ref{sup:manifold} in the supplementary materials for an introduction to the manifold.

\subsection{Online FPCA}

For simplicity, we assume $\mu(\bt)=\E\{X(\bt)\}=0$ and focus on the estimation of FPCs. For the case with a nonzero mean function, we can center each $\by_i$ by subtracting an estimated mean. See Section~\ref{sup:mean} in the supplementary materials for discussion about online mean function estimation.

We first introduce a valid objective function for a single observed subject and then discuss the general case where multiple subjects arrive as mini-batches. For the $i$th subject $X_i(\bt)$, its observed covariance $\bC_i = \by_i \by_i^\top$ is an unbiased estimator for the true covariance $[\gamma(\bt_{ij_1}, \bt_{ij_2})]_{j_1,j_2} + \sigma^2 \bI_{m_i}$, where the nugget term $\sigma^2 \bI_{m_i}$ accounts for the observational noise.
Recall that the covariance function $\gamma$ can be expanded with the eigenfunctions $\phi_r$'s that are approximated by a spline basis. Write $\bB_i$ as the $m_i\times p$ spline basis matrix of $\bb(\bt)$ evaluated at the $m_i$ observation locations. It follows from the rank-$R$ expression \eqref{eq:cov-rankr} that the above true covariance of $\by_i$ can be approximated by
\begin{equation*}
	\begin{split}
		\bSigma_i(\bTheta, \blambda, \sigma^2) = \bB_i \bTheta\, \rmDiag(\blambda)\, \bTheta^\top \bB_i^\top + \sigma^2\bI_{m_i},
	\end{split}
\end{equation*}
where $\blambda=(\lambda_1,\dots,\lambda_R)$ and $\rmDiag(\cdot)$ returns a diagonal matrix from given elements.
Following \cite{peng2009geometric}, we use
\begin{equation}\label{eq:obj-lik}
\ell_i(\bTheta, \blambda, \sigma^2)
= \rmtr\left [ \{\bSigma_i(\bTheta, \blambda, \sigma^2)\}^{-1} \bC_i \right ] + \log\big| \bSigma_i(\bTheta, \blambda, \sigma^2) \big|
\end{equation}
to quantify the discrepancy between the observed covariance $\bC_i$ and the spline approximation model $\bSigma_i(\bTheta, \blambda, \sigma^2)$.
\citet{peng2009geometric} noted that $\ell_i(\bTheta, \blambda, \sigma^2)$ is nonconvex, but is locally convex in a neighborhood of the true parameters.
The above covariance discrepancy corresponds to the negative log-likelihood of $\by_i$ up to some constant under the assumption that $\by_i$ follows a multivariate normal distribution. 
From the perspective of matrix Bregman divergence \citep{dhillon2008matrix,kulis2009low}, the discrepancy measure \eqref{eq:obj-lik} corresponds to the {\it LogDet divergence} or {\it Stein's loss}, generated by a negated logarithm seed function. This flexible family of divergences includes other popular loss functions for covariance fitting, such as the {\it von Neumann divergence} and the {\it Frobenius norm loss}. Therefore, it is straightforward to extend our method to general loss functions when the multivariate Gaussian assumption does not hold. See recent work by \cite{he2024penalized}.

Now, suppose that we receive mini-batches of subjects from the data stream instead of a single observation. Let $\mathcal{I}_k$ denote the index set of the received observations, and let the $k$th mini-batch contain subjects $\{\by_i : i \in \mathcal{I}_k\}$.
It is natural to use the average of the covariance discrepancies plus a roughness penalty as the objective function for model parameter $(\bTheta,\blambda,\sigma^2)$, that is
\begin{equation}\label{eq:fk}
	\begin{split}
		 & f_k(\bTheta, \blambda, \sigma^2; \tau) = \frac{1}{|\mathcal{I}_k|} \sum_{i\in \mathcal{I}_k} \ell_i(\bTheta, \blambda, \sigma^2) + \tau h(\bTheta),
	\end{split}
\end{equation}
where $|\mathcal{I}_k|$ is the size of the mini-batch, and $\tau>0$ is the smoothing parameter balancing the data fidelity and the roughness penalty $h(\bTheta)$. Popular choices of the roughness penalty include tensor-product smoothing penalty and thin-plate regression penalty \citep{wood2017generalized}. More detail on the penalty is in Section~\ref{sup:bspline} in the supplementary materials.

\section{Riemannian Stochastic Gradient Descent}\label{sec:rsgd-fix}

In this section, we develop a Riemannian Stochastic Gradient Descent (RSGD) algorithm to update the model parameters efficiently along the geometric structure. The smoothing parameter $\tau$ is assumed to be fixed for now and will be relaxed with a dynamic tuning strategy in the next section.
 
RSGD is an extension of the standard stochastic gradient descent tailored for manifold-constrained problems. The main idea of RSGD is to evaluate the fit of the current estimate to the newly received mini-batch, compute the Riemannian gradient as the update direction, and project the parameters onto a new position on the manifold. In the following, we explain the updating rules for each component of the model parameter. Denote the updated model parameter at iteration $k$ as $(\bTheta_k,\blambda_k,\sigma^2_k)$.

For the spline coefficients $\bTheta$, the Riemannian gradient at iteration $k$, denoted by $\nabla_{\bTheta} f_{k}$, is obtained via $\nabla_{\bTheta} f_k(\bTheta_{k-1}, \blambda_{k-1}, \sigma^2_{k-1}; \tau) = \calP_{\bTheta_{k-1}} \left( \partial_\bTheta f_k(\bTheta_{k-1}, \blambda_{k-1}, \sigma^2_{k-1}; \tau) \right)$, where the projection operator $\calP_{\bTheta}$ projects the Euclidean gradients $\pp_\bTheta f_k$ onto the tangent space $T_{\bTheta}\stg$. See Section~\ref{sup:manifold} in the supplementary materials for an introduction to projection and tangent space. Intuitively, the projection removes the component of the gradient that is perpendicular to the manifold, resulting in a refined search direction for the manifold-constrained problem. The previous estimate, \(\bTheta_{k-1}\), is then updated along the Riemannian gradient. In RSGD, this update can be performed using either an exponential map or a retraction, both of which serve as manifold-specific analogs of standard gradient descent. Here, we focus on retraction due to its relative simplicity and computational efficiency. The retraction step at iteration $k$ with a step size $\alpha_k$ is written as
\[
	\bTheta_{k} = \calR_{\bTheta_{k-1}}(- \alpha_k \nabla_{\bTheta} f_{k}(\bTheta_{k-1}, \blambda_{k-1}, \sigma^2_{k-1}; \tau)),
\]
where $\calR_{\bTheta}$ is a retraction operator that smoothly maps the tangent space $T_\bTheta \stg$ to the manifold $\stg$ and satisfies $\calR_{\bTheta}(\mathbf{0}_{p\times R})=\bTheta$ and $\frac{\rmd}{\rmd \alpha}\big|_{\alpha=0} \calR_{\bTheta}(-\alpha \bDelta) = \bDelta$ for any $\bTheta \in \stg$ and $\bDelta\in T_{\bTheta} \stg$.

For $\blambda$ and $\sigma^2$, we reparameterize them as $\blambda = \exp(\bmeta) + \delta$ with $\bmeta \in \mathbb{R}^R$, and $\sigma^2 = \exp(\zeta) + \delta$ with $\zeta \in \mathbb{R}$, where $\delta$ is a small positive constant. This approach ensures that $\blambda$ and $\sigma^2$ remain bounded away from zero, thereby guaranteeing the non-singularity of $\bSigma_i(\bTheta, \blambda, \sigma^2)$, while simultaneously avoiding explicit positivity constraints. Since $\bmeta$ and $\zeta$ lie in the entire Euclidean space, their RSGD updates reduce to the standard SGD.
\spacingset{1}
\begin{algorithm}[tb]
	\DontPrintSemicolon
	\caption{Averaged RSGD for online FPCA (OnlineFPCA-RSGD)}
	\label{alg:rsgd}
	\KwInput{$(\by_i, \bt_i)_{i=1}^N$, $(\alpha_k)_{k=1}^\infty$, $(\mathcal{I}_k)_{k=1}^\infty$, $\delta$ }
	\KwInit{$\bTheta_0,\bmeta_0 = \log(\blambda_0 - \delta)$, $\zeta_0=\log(\sigma^2_0 - \delta),\wb\bTheta_0 = \bTheta_0,\wb\bmeta_0=\bmeta_0,\wb\zeta_0=\zeta_0$}
	\For{$k = 1,2,\ldots$}{
	Select smoothing parameter $\tau_k$\;
	\;
	\texttt{/* Calculate Riemannian stochastic gradients */}\;
	$\mathbf{S}_{k}^{\bTheta} = \calP_{\bTheta_{k-1}} \big(\frac{1}{|\mathcal{I}_k|} \sum_{i\in \mathcal{I}_k} \nabla_\bTheta\, \ell_i(\bTheta_{k-1}, \blambda_{k-1}, \sigma_{k-1}^2) + \tau_k \nabla_\bTheta\, h(\bTheta_{k-1}) \big)$\;
	$\mathbf{s}_{k}^{\bmeta} = \frac{1}{|\mathcal{I}_k|} \sum_{i\in \mathcal{I}_k} \nabla_{\bmeta}\, \ell_i(\bTheta_{k-1}, \blambda_{k-1}, \sigma_{k-1}^2)$\;
	$s_{k}^{\zeta} = \frac{1}{|\mathcal{I}_k|} \sum_{i\in \mathcal{I}_k} \nabla_{\zeta}\, \ell_i(\bTheta_{k-1}, \blambda_{k-1}, \sigma_{k-1}^2)$\;
	\;
	\texttt{/* RSGD updates */}\;
	$\bTheta_{k} = \calR_{\bTheta_{k-1}}(- \alpha_k \mathbf{S}_{k}^{\bTheta})$ \;
	$\bmeta_{k} = \bmeta_{k-1} - \alpha_k \mathbf{s}_{k}^{\bmeta}$ \;
	$\zeta_{k} = \zeta_{k-1} - \alpha_k s_k^{\zeta}$\;
	\;
	\texttt{/* Iterative averaging */}\;
	$\widebar\bTheta_{k} = \calR_{\widebar\bTheta_{k-1}}\big(\frac{1}{k}\calR^{-1}_{\widebar\bTheta_{k-1}}(\bTheta_{k})\big)$\;
	$\widebar\bmeta_{k} = \widebar\bmeta_{k-1} + \frac{1}{k} (\bmeta_k - \widebar\bmeta_{k-1})$\;
	$\widebar\zeta_{k} = \widebar\zeta_{k-1} + \frac{1}{k} (\zeta_k - \widebar\zeta_{k-1})$\;
	}
\end{algorithm}
\spacingset{1.8}

To conclude this section, we introduce the asymptotic properties of the RSGD algorithm with a fixed smoothing parameter $\tau$.
By the stochastic approximation theory \citep{bonnabel2013stochastic,bottou2018optimization}, as the number of mini-batches increases, the online updates of the model parameter will converge to a stationary point of the population loss function
\begin{equation}\label{eq:obj-batch}
	F(\bTheta, \blambda, \sigma^2; \tau) = \E \big[ \ell_i(\bTheta, \blambda, \sigma^2)\big] + \tau h(\bTheta),
\end{equation}
where the expectation is taken on $(\bt_i,\by_i)$'s.
Typically, RSGD is sample-inefficient and its convergence rate in nonconvex settings is suboptimal \citep{zhang2016first}. It is commonly used as a compromise between sample efficiency and computational cost in large-data settings \citep{bottou2018optimization}. When repeated access to the full dataset is feasible, additional epochs can make more effective use of the data and typically improve estimation accuracy. 
Crucially, \citet{polyak1964some} and \cite{polyak1992acceleration} found in the Euclidean setting that a simple arithmetic average of SGD iterates achieves optimal parametric convergence rate and asymptotic normality.
This averaging technique was later extended to RSGD by \citet{tripuraneni2018averaging} via defining the Riemannian average. When applied to our RSGD iterates $\bTheta_k$, the averaged iterate sequence is
\[\widebar\bTheta_{0} = \bTheta_{0},\quad \widebar\bTheta_{k} = \calR_{\widebar\bTheta_{k-1}}\bigg(\frac{1}{k}\calR^{-1}_{\widebar\bTheta_{k-1}}(\bTheta_{k})\bigg)  \text{ for } k \ge 1. \]
For the other Euclidean model parameters, the averaged estimates reduce to the regular iterative averaging that starts with $\widebar{\bmeta}_0 = \bmeta_0$ and $\widebar{\zeta}_0 = \zeta_0$, and updates via
\begin{equation*}
	\widebar{\bmeta}_k = \widebar{\bmeta}_{k-1} + \frac{1}{k} (\bmeta_k - \widebar{\bmeta}_{k-1}), \quad
	\widebar{\zeta}_k = \widebar{\zeta}_{k-1} + \frac{1}{k} (\zeta_k - \widebar{\zeta}_{k-1}).
\end{equation*}
Algorithm~\ref{alg:rsgd} displays the complete procedure of RSGD and averaging, despite the smoothing parameter tuning is delayed to the next section.
In Section~\ref{sec:conv}, we will show that both the convergence of RSGD and the asymptotic normality of the averaged RSGD hold for a varying sequence of smoothing parameters. In particular, the asymptotic normality will be used to construct confidence intervals for FPCs in Section~\ref{sup:ci-online} of the supplementary materials.

\section{Dynamic Tuning}\label{sec:dyn-tuning}

In this section, we design a novel dynamic tuning scheme for the smoothing parameter $\tau$ which is previously kept fixed in the objective function \eqref{eq:fk}. In the literature on functional data analysis, smoothing parameter selection is crucial to estimating accuracy and model interpretability. Existing batch methods commonly use cross-validation and its variants to achieve model tuning. However, the multi-fold data split required by cross-validation is not applicable to the online setting where data arrive sequentially. Even when we have access to the full dataset such that cross-validation is possible, the computational cost of searching the best among a fine grid of smoothing parameters is prohibitively high. It thus contradicts the fundamental goal of designing online methods for computational efficiency.

\begin{figure}[!ht]
	\centering
	\includegraphics[width=\linewidth]{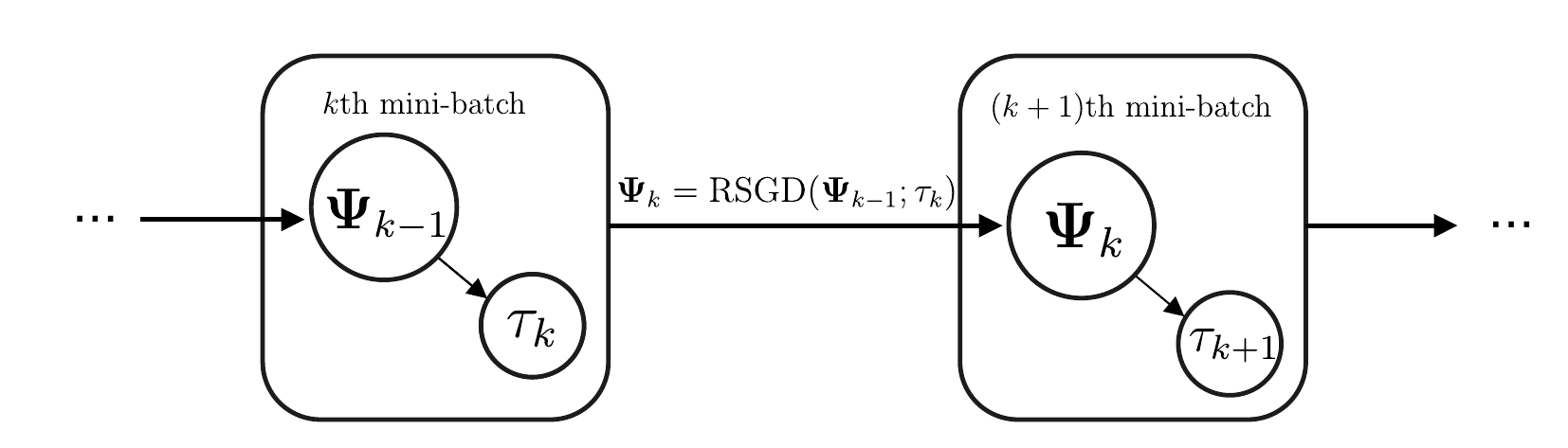}
	\caption{A schematic diagram of the online estimation with dynamic tuning. At the $k$th iteration, a new smoothing parameter $\tau_k$ is chosen using the $k$th mini-batch and the current model parameter estimate $\bPsi_{k-1}$. Then, the model parameter is updated to $\bPsi_{k}$ via RSGD. Consecutively, at the $(k+1)$th iteration, $\tau_{k+1}$ is chosen based on the $(k+1)$th mini-batch and $\bPsi_k$, and the procedure proceeds. The pair $(\bPsi_{k-1},\tau_k)$ is presented in a rectangle as a node, and the RSGD step is a path connecting subsequent nodes. The node-path view motivates us to develop the beam search in Section~\ref{sec:beam-search} to dynamically select $\tau_k$.}
	\label{fig:online-pipeline}
\end{figure}

To address this issue, we dynamically update the smoothing parameter along with the model parameter updating process. Specifically, when the $k$th mini-batch arrives, the RSGD update of model parameter $\bTheta_{k}$ with dynamic tuning $\tau_{k}$ is
\begin{equation*}\label{eq:rsgd-varytau}
	\bTheta_{k} = \mathcal{R}_{\bTheta_{k-1}}\bigg(-\alpha_k \nabla_\bTheta f_{k} (\bTheta_{k-1}, \blambda_{k-1}, \sigma^2_{k-1}; \tau_{k}) \bigg),
\end{equation*}
where the smoothing parameter $\tau_k$ is allowed to change with $k$. Other model parameters that are not subject to smoothness regularization, namely $\blambda$ and $\sigma^2$, are still updated by the same approach as in the previous section.
For simplicity, we write $\bPsi=(\bTheta, \bmeta, \zeta)$, denote its estimate at the $k$th iteration by $\bPsi_k$, and abbreviate the $k$th parameter update as $\bPsi_k = \mathrm{RSGD}(\bPsi_{k-1}; \tau_k)$. Figure~\ref{fig:online-pipeline} displays the pipeline of model parameter update and dynamic tuning.
\spacingset{1}
\spacingset{1.8}
Our dynamic tuning strategy is characterized by two primary components: a goodness-of-fit measure of the current model parameter estimate to the data, and an efficient approach to acquire new smoothing parameter candidates. To these ends, we first propose a novel {\it averaged block validation} (ABV) score, which iteratively combines validation scores from each newly received data block using an exponential weighting scheme.
Second, we develop a beam search algorithm to explore the space of smoothing parameters. While only a few rather than a dense grid of candidate smoothing parameters are maintained during the process, the one with the smallest ABV score is selected.

\subsection{Averaged Block Validation Score}\label{sec:abv}

Recall that we define in Equation~\eqref{eq:obj-lik} the covariance discrepancy measure $\ell_i(\bTheta,\blambda,\sigma^2)$. For brevity, we write the discrepancy measure as $\ell_i(\bPsi)$ with $\bPsi=(\bTheta, \bmeta, \zeta)$ collecting the model parameters.
When a new mini-batch arrives, let
$
\texttt{V}_k = {\sum_{i\in \mathcal{I}_k} \ell_i(\bPsi_{k-1})}/{|\mathcal{I}_k|}
$
be the validation score of $\bPsi_{k-1}$ using the $k$th mini-batch.
We then obtain the {\it averaged validation} (AV) score by accumulating $\texttt{V}_k$'s with a constant $\omega\in(0,1)$
\begin{equation}\label{eq:av-exp}
	\texttt{V}_0 = 0, \qquad \texttt{AV}_{k} = (1 - \omega) \texttt{V}_{k} + \omega \texttt{AV}_{k-1}.
\end{equation}
It is important to note that $\texttt{AV}_{k}$ is a history of past out-of-sample performances. We do not re-evaluate the current model on previous mini-batches. This provides a stable metric of the learning trajectory's generalization error without violating the principle of using only unseen data for validation.
In fact, the AV score can be expressed as the exponentially weighted moving average of $\texttt{V}_k$'s, that is,
$\texttt{AV}_{k} = (1 - \omega) \sum_{j=0}^{k-1} \omega^{j} \texttt{V}_{k-j}.$
It is clear that the weight $\omega^j$ decays fast as $j$ increases. Therefore, the AV score places more importance on the more recent validation performance, which justifies its use in the online setting. 

In our experiments, we found that the AV score often exhibits significant fluctuations across iterations, especially when the mini-batch size is small. A partial explanation is that the score is likely dominated by a few recent mini-batches, due to the rapid decay of the weight $\omega^j$. To stabilize the variation of the AV score, we propose to form several adjacent mini-batches as a data block and introduce the {\it averaged block validation} (ABV) score as follows. First, let $q$ be the number of mini-batches in each data block, hence the index $k$ of the mini-batches in the $n$th data block satisfies $k\in \{(n-1)q + 1, \dots, nq\}$. The {\it block validation} (BV) score of the $n$th data block is defined as the average of the validation scores across all involved mini-batches, i.e., $\texttt{BV}_n = \sum_{k=(n-1)q + 1}^{nq} \texttt{V}_k / q$. Then, we compute the ABV score using the exponential moving average such that
$
\texttt{ABV}_0 = 0, \texttt{ABV}_n = (1-\omega)\texttt{BV}_{n} + \omega \texttt{ABV}_{n-1}.
$
Clearly, the ABV score reduces to the AV score in Equation~\eqref{eq:av-exp} when $q=1$, and to the average of all $\texttt{V}_k$'s as $q\to\infty$. Therefore, the ABV score serves as an intermediate between the AV score and the arithmetic mean of the validation scores. It effectively tracks validation performance throughout the sequential process, providing a more stable metric by incorporating more mini-batches.

\subsection{Beam Search}
\label{sec:beam-search}
\spacingset{1}
\begin{figure}[!ht]
	\centering
	\includegraphics[width=\linewidth]{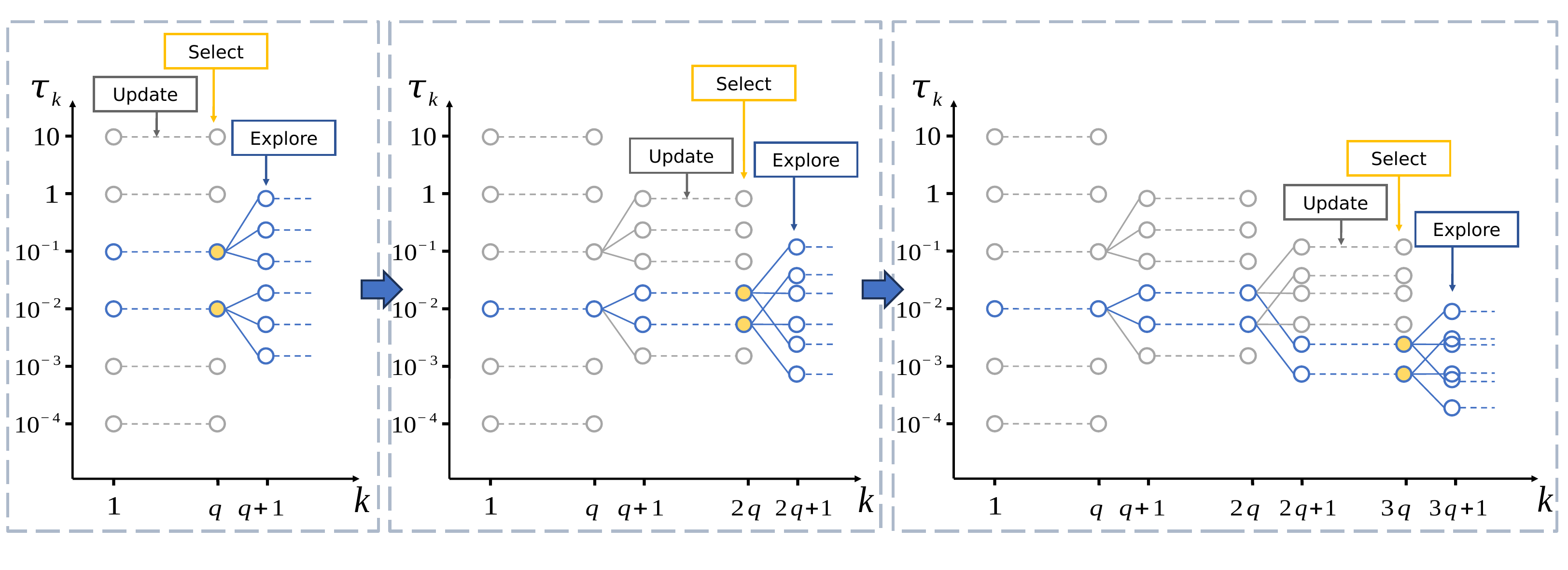}
	\caption{A beam search diagram for the first three rounds of the dynamic tuning. The beam width $W=2$, and the branching factor $B=3$, such that $C=W\times B=6$ candidate smoothing parameters are maintained throughout. Each node represents a smoothing parameter and a model parameter estimate $(\tau_k, \bPsi_{k})$ at iteration $k$. Dashed lines indicate RSGD steps from new to old nodes. During the update stage, model parameters are updated through $q$ RSGD iterations (dashed lines) with smoothing parameters remaining fixed. Then the two nodes (highlighted in yellow) with the lowest ABV scores are selected, and each is further expanded into three new nodes (solid lines). Active paths are displayed in blue, whereas discarded paths are shown in gray.}
	\label{fig:sketch-tuning}
\end{figure}
\spacingset{1.8}

Beam search is a heuristic tree-search algorithm widely used to construct a near-optimal sequence of decisions.
The core idea is to try $C$ candidate decisions, evaluate their performances, keep the $W$-best candidates, then expand each retained candidate into $B$ different decisions in the next step, and repeat the procedures. We refer to $C$ as the number of candidates, $W$ as the beam width, and $B$ as the branching factor. We impose $C = W \times B$ such that the number of candidates each step remains fixed.  

In our dynamic tuning scheme, we formulate the online search for the best smoothing parameter as a tree-search problem and solve it using a beam-search-style procedure.
At each data block, we run RSGD in parallel under $C$ candidate smoothing parameters, we evaluate the resulting parameter estimates using the recorded ABV scores, and retain only the top $W$ candidates.
We then expand each selected candidate by proposing $B$ nearby smoothing parameters for the next data block, giving $C = W \times B$ candidates again.
Figure~\ref{fig:sketch-tuning} provides a schematic illustration, where each node represents a pair consisting of a smoothing parameter and its associated model parameter estimate.

We now detail the three stages of dynamic tuning: update, selection, and exploration. Initialize $C$ smoothing parameters $\{\tau_{0}^{(c)}\}_{c=1}^C$, and let $\{\bPsi_0^{(c)}\}_{c=1}^C$ be $C$ copies of the initial model parameter estimate. Recall that the $n$th data block contains $q$ mini-batches with indices of the mini-batches in $\{(n-1)q + 1, \dots, nq\}$. We perform the following steps.

\noindent\textbf{Stage 1: Update.} For the $n$th data block, i.e., the mini-batch indices $k$ satisfying $(n-1)q + 1 \leq k\leq nq$, and for each $c=1,\dots,C$, we keep the smoothing parameter fixed within the block by setting $\tau_k^{(c)} = \tau_{k-1}^{(c)}$. We then update the model parameters by $\bPsi_k^{(c)} = \mathrm{RSGD}(\bPsi_{k-1}^{(c)}; \tau_{k}^{(c)})$, and accumulate associated BV scores, denoted by $\{\texttt{BV}_n^{(c)}\}_{c=1}^{C}$.

\noindent\textbf{Stage 2: Selection.} After finishing the $n$th data block (i.e., at iteration $k_n = nq$), we update the ABV scores $\texttt{ABV}_n^{(c)}$ by averaging newly accumulated $\texttt{BV}_n^{(c)}$ into $\texttt{ABV}_{n-1}^{(c)}$. We then select $W$ candidates with the smallest ABV scores. Let the selected indices be $c_{n1}, \dots, c_{nW}$, so the chosen estimates and their associated smoothing parameters are $\{\bPsi_{k_n}^{(c_{nw})}\}_{w=1}^W$ and $\{\tau_{k_n}^{(c_{nw})}\}_{w=1}^W$, respectively.

\noindent\textbf{Stage 3: Exploration.} We explore the smoothing parameter space by proposing $B$ new smoothing parameters in a neighborhood of each selected $\tau_{k_n}^{(c_{nw})}$, which sum up to $C$ new candidates, denoted by $\{\tau_{k_n+1}^{(c)}\}_{c=1}^{C}$. If for some $c\in\{1,\dots,C\}$ and $w\in\{1,\dots,W\}$, $\tau_{k_n+1}^{(c)}$ is from the neighborhood of $\tau_{k_n}^{(c_{nw})}$, we require that, for some $\beta > 1$ and some $A_1,A_2>0$,
\begin{equation}\label{eq:logtauk}
\tau_{k_n+1}^{(c)} \in \Big[\tau_{k_n}^{(c_{nw})} e^{-A_1 n^{-\beta}},\ \tau_{k_n}^{(c_{nw})} e^{ A_2 n^{-\beta}}\Big].
\end{equation}
This condition places $\log(\tau_{k_n+1}^{(c)})$ in a neighborhood of $\log(\tau_{k_n}^{(c_{nw})})$ with decaying widths, which guarantees the positivity of smoothing parameters and encourages the tuning to stabilize over time. 
We then return to Stage~1 for data block $n+1$.

The update-select-explore schema identifies paths through the beams-search tree: there exists an index sequence $\{c_k\}_{k\ge0}$ induced by the dynamic tuning procedure such that
\[
\bPsi_k^{(c_k)}=\mathrm{RSGD}\big(\bPsi_{k-1}^{(c_{k-1})};\tau_k^{(c_k)}\big),\qquad k=1,\dots,\infty.
\]
We will study the convergence of the selected $\{\bPsi_k^{(c_k)}\}_{k\ge 1}$ in the next section.
Lemma~\ref{lem:difflogtau} in the supplementary materials shows that the resulting sequence of smoothing parameters satisfies 
$|\tau_k^{(c_k)}-\tau_{k-1}^{(c_{k-1})}|\lesssim k^{-\beta}$ for the same $\beta>1$ as in Equation~\eqref{eq:logtauk}. Thus, the sequence $\{\tau_k\}$ has finite variation, i.e., $\sum_{k=1}^\infty |\tau_k^{(c_k)}-\tau_{k-1}^{(c_{k-1})}| < \infty$, which matches the requirement of Theorem~\ref{thm:conv}. In addition to Equation~\eqref{eq:logtauk}, there exist alternative exploration proposals that are both practical and compatible with the convergence analysis. See Remark~\ref{rmk:altn-explore} in Section~\ref{sup:ass} of the supplementary materials for further discussion.

Both the beam width $W$ and branching factor $B$ are critical to the performance. Larger values of $W$ or $B$ enable more comprehensive exploration of the parameter space, but increase computational cost. In Section~\ref{sup:set-cwb} of the supplementary materials, we compare several $(C,W,B)$ configuration  and show that even a small number of candidates (e.g., $C=3,4,6$) can perform well in practice.
Overall, the beam-search strategy substantially reduces computation relative to exhaustive search while still effectively exploring the parameter space. The trade-off is that, as a heuristic method, it does not guarantee globally optimal tuning; instead, it aims to achieve a favorable balance between efficiency and accuracy.

\section{Convergence Analysis}\label{sec:conv}

We prove the convergence of our method based on the standard RSGD theory under mild conditions. The dynamic tuning brings additional challenges, since the objective function varies throughout the online procedure.

Recall the shorthand notation $\bPsi = (\bTheta, \bmeta, \zeta)$, where $\bTheta$ is the basis coefficient matrix and $\bmeta$ and $\zeta$ are the logarithmic transformations of $\blambda$ and $\sigma^2$, respectively. The parameter space is the product manifold $\calM = \stg \times \mathbb{R}^{R+1}$ that accommodates all $\bPsi$, and $\nabla$ denotes the Riemannian gradient operator on $\calM$ with respect to $\bPsi$. With a slight abuse of notation, we do not distinguish between evaluating a function at $\bPsi$ and at the corresponding tuple $(\bTheta, \blambda, \sigma^2)$.
The objective function $F(\bPsi;\tau)$, defined in Equation~\eqref{eq:obj-batch}, is the expected covariance discrepancy plus a roughness penalty.

\begin{assumption}\label{as:bounded-iter}
The RSGD iterates $\{\bPsi_k\}_{k\geq 1}$ are contained in a compact subset of $\calM$.
\end{assumption}
\begin{assumption}
\label{as:2nd-gradmnt} There exists a constant $M_2 > 0$ such that
$\E[\|\nabla l_i(\bPsi_{k})\|_2^2] \leq M_2$ for all subject $i$ and iteration $k \geq 1$.
\end{assumption}
\begin{assumption}
\label{as:4th-gradmnt} There exists a constant $M_4 > 0$ such that
$\E[\|\nabla l_i(\bPsi_{k})\|_2^4] \leq M_4$ for all subject $i$ and iteration $k \geq 1$.
\end{assumption}
\begin{assumption}
\label{as:eiggap}
The $R+1$ largest eigenvalues satisfy $\lambda_1 > \dots > \lambda_R > \lambda_{R+1}\geq\dots\geq 0$.
\end{assumption}
Assumptions~\ref{as:bounded-iter}--\ref{as:eiggap} are used in the following results. Detailed discussion of these conditions is provided in Section~\ref{sup:ass} of the supplementary materials. 
In summary, Assumption~\ref{as:bounded-iter} (bounded iterations) generally holds in practice or can be enforced by parameter clipping; Assumptions~\ref{as:2nd-gradmnt} and \ref{as:4th-gradmnt} are implied by bounded moments of the observed vectors $\by_i$, together with bounded iterates and a bounded number of measurements per subject. 
In real data applications, these population moment conditions cannot be verified exactly from a finite dataset. We therefore assess their practical plausibility by monitoring the stability of the online updating path, including the selected smoothing parameters, estimated model parameters, and empirical moments of stochastic gradient norms across data blocks. Assumption~\ref{as:eiggap} (positive eigengap) is a natural condition in FPCA studies \citep{paul2009consistency,peng2009geometric} that ensures identifiability of the leading FPCs.

\begin{theorem}\label{thm:conv}
Suppose Assumptions~\ref{as:bounded-iter}--\ref{as:2nd-gradmnt} hold. Let $\{\bPsi_{k}\}_{k\geq1}$ be the RSGD iterates and $\{\tau_{k}\}_{k\geq1}$ be  the associated smoothing parameters yielded by the dynamic tuning scheme. Suppose that the step sizes $\{\alpha_k\}_{k\geq 1}$ satisfy $ \sum_{k=1}^\infty \alpha_k = \infty$ and $\sum_{k=1}^\infty \alpha_k^2 < \infty$, and the smoothing parameters $\{\tau_{k}\}_{k\geq1}$ have finite total variation, i.e., satisfies $\sum_{k=1}^\infty |\tau_{k+1} - \tau_k| < \infty$.
Then, $\lim_{k\to\infty}\tau_k = \tau_\star$ for some $\tau_\star \geq 0$, and $\{\bPsi_k\}_{k\geq1}$ converges almost surely to a stationary point on $\calM$ in the sense that the Riemannian gradient norm vanishes asymptotically:
\[
    \| \nabla F(\bPsi_{k};\tau_\star)\|^2 \to 0 \quad \text{a.s.} \qquad \text{as $k\to \infty$}.
\]
\end{theorem}
Proof of Theorem~\ref{thm:conv} is in Section~\ref{sup:conv} of the supplementary materials.
Theorem~\ref{thm:conv} describes the convergence of the RSGD algorithm with a dynamic smoothing parameter. The requirements and assumptions are relatively mild: The diminishing step-size scheme is commonly used in stochastic approximation algorithms to ensure convergence \citep{bottou2018optimization}; the finite total variation of $\{\tau_k\}_{k\geq1}$ is naturally satisfied by using an appropriate exploration scheme such as Equation~\eqref{eq:logtauk} in Section~\ref{sec:beam-search} or Equation~\eqref{eq:taukbeta} in Section~\ref{sup:ass} of the supplementary materials.

\noindent{\bf Remark:} It is important to distinguish between the convergence of the optimization algorithm and the statistical consistency of the estimator. Theorem~1 guarantees that the RSGD iterates converge to a stationary point of the objective function. For statistical convergence to the true eigenfunctions, one would require $\tau_k \to 0$.
However, characterizing the asymptotic behavior for penalized tensor-product splines on multidimensional domains remains an open problem, even in the classical batch setting. The analysis is further complicated in the online setting by the interplay between the stochastic gradient noise and the changing objective geometry. A formal derivation of the asymptotic statistical properties is left for future research.

Our next theorem shows that the averaged RSGD iterates with dynamic smoothing parameter tuning can reach the same asymptotic distribution as that given in \citet{tripuraneni2018averaging} whose objective function has no tuning parameter. Let $\bPsi_\star = \lim_{k\to\infty} \bPsi_k$ be the limiting estimate and $\wb\bPsi_k$ be the iteratively averaged estimates as defined in Section~\ref{sec:rsgd-fix}. We denote $\calH_{\tau_\star}: \calT_{\bPsi_\star}\calM \to \calT_{\bPsi_\star}\calM$ the Hessian operator of $F(\cdot;\tau_\star)$ at $\bPsi_\star$. The detailed definition of the Riemannian Hessian operator is provided in Section~\ref{sup:hess} in the supplementary materials. The inverse $\calH_{\tau_\star}^{-1}$ is understood on the tangent space at the limiting stationary point. Define $\calS_{\tau_\star} = \E[\nabla f_k(\bPsi_{\star}; \tau_\star) \otimes \nabla f_k(\bPsi_{\star}; \tau_\star)]$, the covariance of the Riemannian gradient of the individual loss function at the stationary point $\bPsi_{\star}$.

\begin{theorem}\label{thm:normality}
Suppose Assumptions~\ref{as:bounded-iter}, \ref{as:4th-gradmnt}, and \ref{as:eiggap} hold. Let $\{\bPsi_k\}_{k\geq1}$ be the RSGD iterates with step sizes $\alpha_k \asymp k^{-\alpha}$ for $1/2 < \alpha < 1$, and with dynamic tuning satisfying $|\tau_{k+1}-\tau_k| = O\big(k^{-(3-\alpha)/2-\epsilon}\big)$ for some $\epsilon > 0$. Then, as $k\to\infty$, the tangent element $\bU_k = \calR_{\bPsi_{\star}}^{-1}(\wb\bPsi_k) \in \calT_{\bPsi_\star}\calM$ satisfies
$\sqrt{k} \bU_k \xrightarrow{d} \calN\big(\mathbf{0}, \calH_{\tau_\star}^{-1} \calS_{\tau_\star} \calH_{\tau_\star}^{-1}\big)$.
\end{theorem}
In Theorem~\ref{thm:normality}, the condition on the smoothing parameters $\{\tau_k\}_{k\geq1}$ can be satisfied by setting $\beta = (3-\alpha)/2+\epsilon$ in Equation~\eqref{eq:logtauk} of the dynamic tuning scheme and then using the technique in Lemma~\ref{lem:difflogtau} to transfer the bound from $|\log(\tau_{k+1}) - \log(\tau_{k})|$ to $|\tau_{k+1} - \tau_{k}|$. 
Moreover, Theorem~\ref{thm:normality} can be used for uncertainty quantification. Section~\ref{sup:ci-online} provides further details on confidence band construction for FPC estimates.

\section{Riemannian AdaGrad for Online FPCA}\label{sec:adagrad}

\spacingset{1}
\begin{algorithm}[tb]
	\DontPrintSemicolon
	\caption{Averaged Riemannian AdaGrad (OnlineFPCA-RAdaGrad)}
	\label{alg:radagrad}
	\KwInput{$(\by_i, \bt_i)_{i=1}^N$, $(\alpha_k)_{k=1}^\infty$, $(\mathcal{I}_k)_{k=1}^\infty$, $\delta$, $0<\beta_1,\beta_2<1$ }
	\KwInit{$\bTheta_0$, $\bmeta_0 = \log(\blambda_0-\delta)$, $\zeta_0=\log(\sigma^2_0-\delta)$,$\wb\bTheta_0 = \bTheta_0,\wb\bmeta_0=\bmeta_0,\wb\zeta_0=\zeta_0$, $\bV_0^{\bTheta}=\mathbf{0}_{R\times R}$, $\bv_0^{\bmeta}=\mathbf{0}_R$, $v_0^\zeta=0$}
	\For{$k = 1,2,\ldots$}{
	Select smoothing parameter $\tau_k$\;
	Compute stochastic gradients $\mathbf{S}_k^\bTheta,\mathbf{s}_k^{\bmeta},s_k^{\zeta}$ as in Algorithm~\ref{alg:rsgd}\;
	\;
	\texttt{/* Accumulate squared gradients */}\;
	$\bV_k^{\bTheta} = \frac{1}{k} \rmDiag(\mathbf{S}_k^\bTheta \bG \mathbf{S}_k^\bTheta)  + \frac{k-1}{k} \bV_{k-1}^{\bTheta}$\;
	$\bv_k^{\bmeta} = \frac{1}{k} (\mathbf{s}_k^{\bmeta})^2 + \frac{k-1}{k} \bv_{k-1}^{\bmeta} $\;
	$v_k^{\zeta} = \frac{1}{k} (s_k^{\zeta})^2 + \frac{k-1}{k} v_{k-1}^{\zeta} $\;
	\;
	\texttt{/* AdaGrad updates */}\;
	$\bTheta_{k} = \calR_{\bTheta_{k-1}}(- \alpha_k \calP_{\bTheta_{k-1}}( \mathbf{S}_k^{\bTheta} {(\bV_k^{\bTheta})^{-1/2}}))$ \;
	$\bmeta_k = \bmeta_{k-1} - \alpha_k \mathbf{s}_k^{\bmeta} \odot (\bv_k^{\bmeta})^{-1/2}$\;
	$\zeta_k = \zeta_{k-1} - \alpha_k s_k^{\zeta} (v_k^{\zeta})^{-1/2}$\;
	\;
	\texttt{/* Iterative averaging */}\;
	$\widebar\bTheta_{k} = \calR_{\widebar\bTheta_{k-1}}\big(\frac{1}{k}\calR^{-1}_{\widebar\bTheta_{k-1}}(\bTheta_{k})\big)$\;
	$\widebar\bmeta_{k} = \widebar\bmeta_{k-1} + \frac{1}{k} (\bmeta_k - \widebar\bmeta_{k-1})$\;
	$\widebar\zeta_{k} = \widebar\zeta_{k-1} + \frac{1}{k} (\zeta_k - \widebar\zeta_{k-1})$\;
	}
	{\textsuperscript{*}}{\footnotesize Notation $\odot$ denotes entry-wise multiplication. Exponents are also entry-wise.}
\end{algorithm}
\spacingset{1.8}
In this section, we extend the adaptive gradient (AdaGrad) algorithm \citep{duchi2011adaptive} to the Riemannian manifold setting and incorporate it as a drop-in replacement for the RSGD step within the online FPCA framework. AdaGrad is a widely used optimization algorithm in machine learning that assigns larger step sizes for parameters with smaller historical gradients. In the RSGD for our online FPCA problem, the step sizes for all FPCs are the same, which may lead to unbalanced convergence speed--relatively fast convergence in some FPCs, typically the leading ones associated with larger eigenvalues, but slow convergence in others. Therefore, an adaptive step size scheme is appealing to balance the convergence across all FPCs.

We present the Riemannian AdaGrad algorithm in Algorithm~\ref{alg:radagrad} for online FPCA. Relative to the standard Euclidean AdaGrad, our procedure has three key adaptations.
First, for the FPC coefficient matrix $\bTheta$, we apply adaptivity to columns rather than individual entries. Let $\bS_j^{\bTheta} = \nabla_{\bTheta} f_j(\bPsi_{j-1}; \tau_j)$ denote the update direction for $\bTheta$ at iteration $j$, and let $\bs_{jr}^{\bTheta}$ be the $r$th column of $\bS_j^{\bTheta}$. For each $r$ and iteration $k$, we accumulate the average quadratic form
$
k^{-1}\sum_{j=1}^k \{\bs_{jr}^{\bTheta}\}^\top \bG \bs_{jr}^{\bTheta}
$
and rescale $\bs_{kr}^{\bTheta}$ by the inverse square root of this quantity to get the adapted updating direction. Using
$
\{\bs_{jr}^{\bTheta}\}^\top \bG \bs_{jr}^{\bTheta}
= \int_{\calT} \big[\{\bs_{jr}^{\bTheta}\}^\top \bb(\bt)\big]^2 \,\rmd \bt$,
this can be interpreted as defining a functional update direction
$u_{jr}(\bt) = \{\bs_{jr}^{\bTheta}\}^\top \bb(\bt)$ for the $r$th FPC $\phi_r(\bt)$ and accumulating
$\int_{\calT} u_{jr}(\bt)^2 \,\rmd \bt$ as its per-component second moment. Thus, unlike the standard AdaGrad, which rescales each scalar entry of $\bS_j^{\bTheta}$ separately, our scheme respects the functional nature of the FPCs and assigns a single adaptive step size to each component. For the Euclidean parameters $\bmeta$ and $\zeta$, we follow the usual AdaGrad strategy by accumulating entrywise second moments of their update directions.
Second, the update for $\bTheta$ must respect its manifold constraint $\bTheta \in \stg$. Accordingly, after rescaling we project the update direction onto the tangent space $\calT_{\bTheta_{k-1}}\stg$ and then apply a retraction to map the iterate back onto the manifold. Lastly, we perform iterative averaging on the manifold for $\bTheta$ as in Algorithm~\ref{alg:rsgd}. Although the averaging is not adequately studied for AdaGrad, we find it very effective to stabilize the estimates in practice.

Finally, we remark on the theoretical properties of this adaptive variant. While the convergence and asymptotic properties established in Theorems~\ref{thm:conv} and \ref{thm:normality} rigorously cover the averaged RSGD with dynamic tuning, extending these guarantees to the adaptive Riemannian setting presents unique theoretical challenges. Existing analyses of adaptive Riemannian methods, for example, \cite{becigneul2018riemannian} and \cite{kasai2019riemannian}, typically rely on assumptions such as geodesic convexity or specific product-manifold structures that do not directly accommodate our nonconvex penalized objective on a generalized Stiefel manifold. Consequently, we present OnlineFPCA-RAdaGrad as a practically motivated extension. As demonstrated in the following simulation studies, this adaptive approach consistently yields superior empirical performance.

\section{Simulation Studies}\label{sec:sim}
We conduct one- and two-dimensional simulation studies to evaluate FPC estimation by the propsed methods, namely OnlineFPCA-RSGD (Algorithm~\ref{alg:rsgd}) and OnlineFPCA-RAdaGrad (Algorithm~\ref{alg:radagrad}), both combined with dynamic tuning. The goal is to compare estimation accuracy and computational efficiency against existing online and batch FPCA methods. The data are generated from the one-dimensional setting of \citet{yang2021online} and the two-dimensional setting of \citet{wang2022low}; full simulation details are provided in Section~\ref{sup:expr-details} of the supplementary materials. In each setting, $N=5000$ noisy irregularly observed functions are generated, the mini-batch size is fixed at 5, and results are averaged over 100 replications.

For the one-dimensional case, we compare with OnlineCov \citep{yang2021online} and the batch methods FACE \citep{xiao2018fast}, REML \citep{peng2009geometric}, and SOAP \citep{nie2022recovering}. For the two-dimensional case, we compare with SOAP-2D \citet{shi2022two}; mOpCov \citet{wang2022low} is not included in the final comparison because it does not scale to the sample size considered here due to memory requirements. OnlineFPCA is initialized using a small subset of observations and then run for multiple epochs. Implementation details, including initialization, burn-in, step-size choices, spline specifications, and dynamic-tuning settings, are given in Section~\ref{sup:expr-details}.

\spacingset{1}
\begin{table}[tb]
\centering
\caption{Running times of online and batch FPCA methods in the simulation studies. OnlineFPCA is run for multiple epochs for better performance. The running times of the OnlineFPCA-RSGD and the OnlineFPCA-RAdaGrad, whose difference is less than 3 seconds in the experiments, are averaged to represent the overall OnlineFPCA method. REML, FACE, and SOAP are the batch methods compared in the 1D case, while SOAP-2D is the only competing method in the 2D case.}
\label{tab:sim-time}
{\small
\begin{tabular}{c c ll S}
\toprule
\multirow{2}{*}{\textbf{Setting}} & \multirow{2}{*}{\textbf{Category}} & \multicolumn{2}{c}{\multirow{2}{*}{\textbf{Method}}} & {\multirow{2}{*}{\textbf{Time (s)}}} \\[-3pt]
  &&  &  &   \\[-3pt]
\midrule
\multirow{7}{*}{1D}   & \multirow{4}{*}{Online}& \multirow{3}{*}{OnlineFPCA}  & 1 epoch  & 28.7  \\[-3pt]
  &&  & 2 epochs & 35.3  \\[-3pt]
  &&  & 3 epochs & 41.1  \\[-3pt]
\cmidrule{3-4}
  && \multicolumn{2}{c}{OnlineCov}& 222.7\\[-3pt]
\cmidrule{2-4}
  & \multirow{3}{*}{Batch} & \multicolumn{2}{c}{REML} & 163.5\\[-3pt]
  && \multicolumn{2}{c}{FACE} & 63.5\\[-3pt]
  && \multicolumn{2}{c}{SOAP} & 114.3\\[-3pt]
\cmidrule{1-5}
\multirow{4}{*}{2D}   & \multirow{3}{*}{Online}& \multirow{3}{*}{OnlineFPCA}  & 1 epoch  & 37.4 \\[-3pt]
  &&  & 3 epochs & 49.4 \\[-3pt]
  &&  & 5 epochs & 61.6 \\[-3pt]
\cmidrule{2-4}
  & Batch  & \multicolumn{2}{c}{SOAP-2D}  & 291.2\\[-3pt]
\bottomrule
\end{tabular}
}
\end{table}

\spacingset{1}
\begin{table}[!htb]
\caption{Estimation accuracy of the top three FPCs by online and batch methods in the simulation studies.
Our two OnlineFPCA algorithms, RSGD and RAdaGrad, are run for multiple epochs for better performance. The lowest RMSE values among online methods are highlighted in bold, with the overall lowest RMSE across all methods marked by an asterisk (``$\ast$''). REML, FACE, and SOAP are the batch methods compared in the 1D case, while SOAP-2D is the only competing method in the 2D case.}
\label{tab:sim-acc}
\centering
{\small
\begin{tabular}{c c ll SSS}
\toprule
\multirow{2}{*}{\textbf{Setting}} & \multirow{2}{*}{\textbf{Category}} & \multicolumn{2}{c}{\multirow{2}{*}{\textbf{Method}}} & \multicolumn{3}{c}{\textbf{RMSE}}  \\[-3pt]
\cmidrule{5-7}
  &&  &   & {\textrm{$\phi_1$}} & {\textrm{$\phi_2$}}  & {\textrm{$\phi_3$}} \\[-3pt]
\midrule
\multirow{10}{*}{1D}   & \multirow{7}{*}{Online}& \multirow{3}{*}{\makecell{OnlineFPCA-   \\[-3pt]RSGD}} & 1 epoch & 0.055 & 0.213 & 0.291 \\[-3pt]
  &&  & 2 epochs  & 0.027 & 0.117 & 0.196   \\[-3pt]
  &&  & 3 epochs  & 0.026 & 0.093 & 0.175   \\[-3pt]
\cmidrule{3-4}
  && \multirow{3}{*}{\makecell{OnlineFPCA-   \\[-3pt]RAdaGrad}} & 1 epoch & 0.049 & 0.187 & 0.270 \\[-3pt]
  &&  & 2 epochs  & 0.024 & 0.092 & 0.167   \\[-3pt]
  &&  & 3 epochs  & \B 0.023 & \B 0.077 & \B 0.153   \\[-3pt]
\cmidrule{3-4}
  && \multicolumn{2}{c}{OnlineCov}& 0.027 & 0.085 & 0.182  \\[-3pt]
\cmidrule{2-4}
  & \multirow{2}{*}{Batch} & \multicolumn{2}{c}{REML} & 0.018 \textsuperscript{*} & 0.057 & 0.127 \\[-3pt]
  && \multicolumn{2}{c}{FACE} & 0.026 & 0.069 & 0.139  \\[-3pt]
  && \multicolumn{2}{c}{SOAP} & 0.042 & 0.068 \textsuperscript{*} & 0.085 \textsuperscript{*}  \\[-3pt]
\cmidrule{1-7}
\multirow{7}{*}{2D}   & \multirow{6}{*}{Online}& \multirow{3}{*}{\makecell{OnlineFPCA-   \\[-3pt]RSGD}} & 1 epoch & 0.116 & 0.167 & 0.178  \\[-3pt]
  &&  & 3 epochs  & 0.071 & 0.101 & 0.113  \\[-3pt]
  &&  & 5 epochs  & 0.066 & 0.095 & 0.105  \\[-3pt]
\cmidrule{3-4}
  && \multirow{3}{*}{\makecell{OnlineFPCA-   \\[-3pt]RAdaGrad}} & 1 epoch & 0.111 & 0.171 & 0.161  \\[-3pt]
  &&  & 3 epochs  & 0.062 & 0.098 & 0.099  \\[-3pt]
  &&  & 5 epochs  & \B 0.054 \textsuperscript{*} & \B 0.083 & \B 0.088 \textsuperscript{*}  \\[-3pt]
\cmidrule{2-4}
  & Batch  & \multicolumn{2}{c}{SOAP-2D}  & 0.054 \textsuperscript{*} & 0.061 \textsuperscript{*} & 0.104  \\[-3pt]
\bottomrule
\end{tabular}
}
\end{table}
\spacingset{1.8}
Table~\ref{tab:sim-time} presents the computational times for both online and batch methods in the simulation examples. OnlineFPCA is substantially faster than the competing methods in these simulations. Despite running for multiple epochs, it completes in significantly less time than competing methods require for a single epoch. Note that the initialization time for OnlineFPCA is included in the first epoch. This efficiency arises from several factors: the direct representation of FPCs rather than relying on the full covariance matrix, the low computational complexity of the RSGD and RAdaGrad algorithms, and an efficient tuning strategy that minimizes redundant computations. Furthermore, the time required for subsequent epochs is significantly shorter than the first, because initialization and dynamic tuning are only included in the first epoch.

Table~\ref{tab:sim-acc} displays the estimation accuracy for the leading three FPCs. As shown by the results, OnlineFPCA methods provide accurate FPC estimations. In the 1D scenario, OnlineFPCA-RSGD performs comparably to the online method OnlineCov, while OnlineFPCA-RAdaGrad achieves accuracy comparable to the best batch method for the first two FPCs. In the 2D scenario, both OnlineFPCA approaches yield accuracy on par with the batch method SOAP-2D. In conclusion, our methods are not only competitive in online FPCA for functional data but also provide competitive alternatives when batch computation is costly.

Figure~\ref{fig:fpca1d-taupath} in the supplementary materials illustrates the dynamic tuning process of one experiment for the 1D simulation study, and a similar plot for the 2D case is included in Section~\ref{sup:extra-figs} of the supplementary materials. In both cases, the dynamic tuning algorithm effectively explores the smoothing parameter space. The selected $\tau_k$ begins at a relatively large value and gradually converges to 0 as $k$ increases and more data become available. This decreasing trend in $\tau_k$ aligns with standard spline smoothing theory, which suggests that the optimal smoothing parameter decreases as the sample size grows.

\section{Real Data Analysis}\label{sec:data}

We illustrate the practical utility and scalability of the proposed OnlineFPCA framework through two applications: spatial air quality monitoring and longitudinal kidney transplant recovery. These datasets illustrate settings where online learning is practically useful because full batch refitting or covariance-based estimation is computationally burdensome.
The first example involves two-dimensional functional data where the curse of dimensionality renders batch covariance tensor estimation computationally prohibitive. The second example involves a massive cohort of subjects. Online learning allows for incremental model updates as new data becomes available, avoiding the need to retrain from scratch. In both cases, our method processes the data sequentially, addressing computational challenges under the resource constraints associated with analyzing the entire dataset at once.
An additional real data analysis of the online news click dataset from Yang and Yao (2023) is provided in Section~S9.8 of the supplementary materials; in that example, OnlineFPCA produces FPC estimates close to OnlineCov as the stream grows while reducing the computing time from 61,493.11 minutes to 96.15 minutes.

\subsection{PM10 Air Quality}
In this section, we study the spatial patterns of the air quality data from the Air Quality System (AQS) (\hyperlink{www.epa.gov/aqs}{https://www.epa.gov/aqs}), an important repository maintained by the United States Environmental Protection Agency (EPA) that takes into account multiple types of pollution, including particulate matter, ground-level ozone, sulfur dioxide, nitrogen dioxide, and carbon monoxide. Specifically, we focus on PM10, which describes inhalable particles with diameters of 10 micrometers or less. The levels of PM10 pollution in the database are recorded as integer indexes called the Air Quality Index (AQI), which are converted from the PM10 concentration. We retrieve the daily AQI from 1982 to 2022 collected at 3,443 monitoring sites across the contiguous United States (the US territory excluding the only two non-contiguous states, Alaska and Hawaii, and all other offshore insular areas). On each day, the data may be only available at a portion of the sites, as different sites record the air quality data intermittently at different frequencies. Figure~\ref{fig:aqs-sites-binned} in the supplementary materials displays the spatial maps of monitoring sites that collected PM10 AQI data on two sampled dates, January 1 and October 31, 2022, showing the irregularities of the data.

In this study, we conduct data analysis by modeling the daily AQI spatial maps as two-dimensional functional data. To preprocess the data, we first transform AQI to its logarithmic value to correct the observable right-skewness. Next, we subtract the mean function, which is estimated by tensor product spline smoothing implemented in the function $\texttt{bam}$ from the R package $\texttt{mgcv}$. Then, we keep only the records on Saturday to reduce temporal dependency and then view the daily measurements as independent functional data. Finally, we bin the monitoring station sites into $2^\circ \times 2^\circ$ boxes and average measurements within each box, creating an aggregated measurement placed at the box center. It avoids repeated computations over the area with extremely dense measurements. After all pre-processing steps, we end up with 2,045 days and a total of 179,619 observations.

We apply the OnlineFPCA-RAdaGrad to extract the leading FPCs of the large and irregularly spaced AQI data. We define tensor product B-spline bases on the box domain $[125^\circ \text{W}, 67^\circ \text{W}] \times [25^\circ \text{N}, 49^\circ \text{N}]$ that contains the contiguous United States, where $8$ marginal basis functions are placed along longitudes and another 6 basis functions along latitudes. The model parameters are initialized by mOpCov with 10\% of the data sampled each day in the first $100$ days. The subsampling used in the initialization is necessary for mOpCov to work well.
OnlineFPCA-RAdaGrad runs with mini-batches of size 10. In dynamic tuning configured by $C=6$, $B=3$, and $W=2$, the smoothing parameter is dynamically selected from three candidates, with initialization $\{10^{-4},\dots,10^{-9}\}$, and is updated for each data block containing 100 days.
Finally, only $50.7$ seconds are used to complete the RAdaGrad procedure. In comparison, the batch method SOAP-2D uses $8.1$ minutes, and mOpCov was not able to complete calculations due to excessive memory requirements by a large multidimensional covariance tensor.

\spacingset{1}
\begin{figure}[tb]
	\centering
	\includegraphics[width=\linewidth]{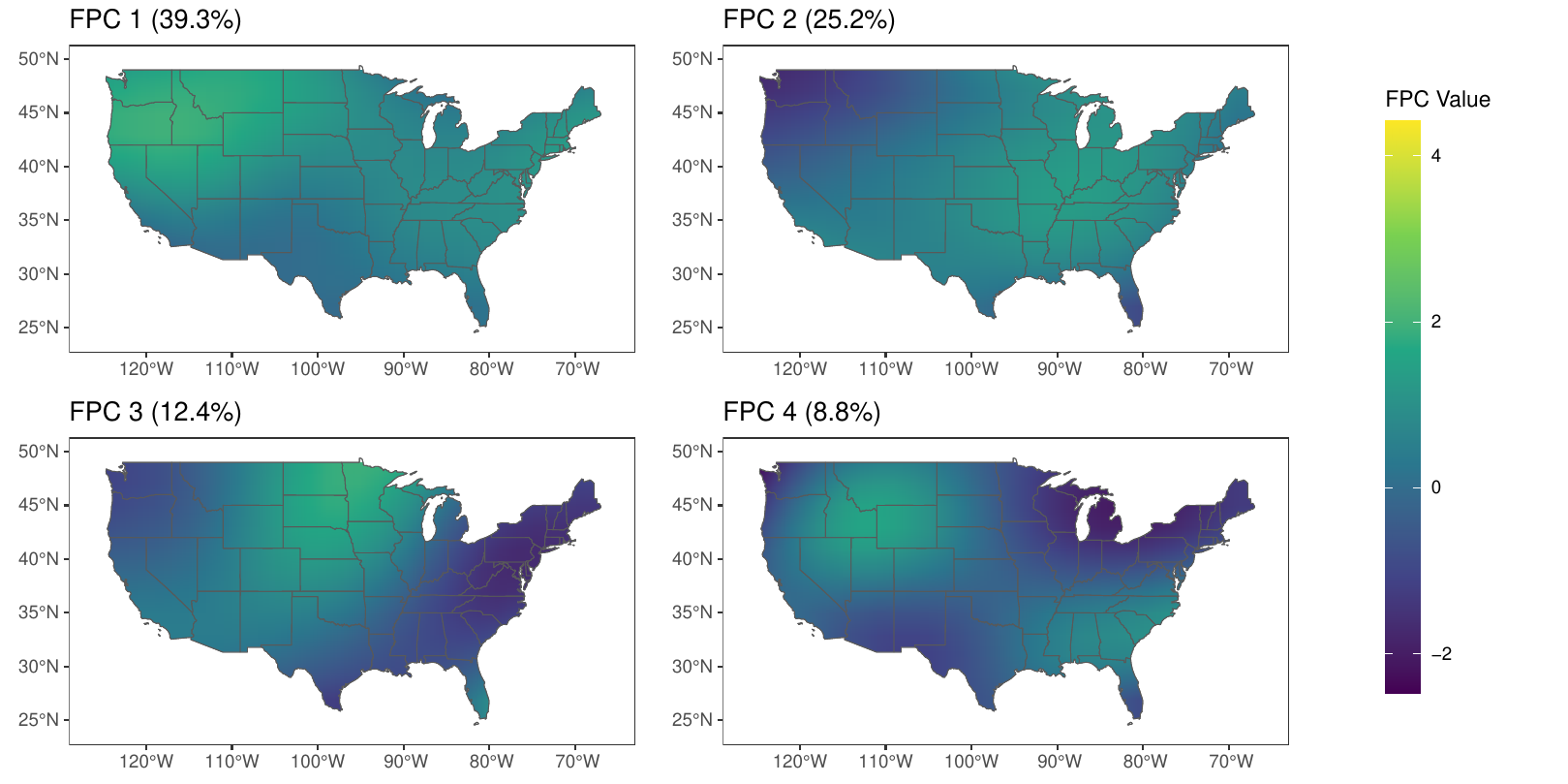}
	\caption{Top four FPCs estimated by OnlineFPCA-RAdaGrad in the air pollution data analysis.}
	\label{fig:aqi-eigfun}
\end{figure}
\spacingset{1.8}
Figure~\ref{fig:aqi-eigfun} presents the first four estimated functional principal components (FPCs). The proportion of variation explained by each FPC, given in parentheses, is determined by the ratio of its estimated eigenvalue $\lambda_k$ to the sum of all eigenvalues. The first FPC (39.3\%) is positive across most of the domain, accounting for a ``common mode'' nationwide. The second FPC (25.2\%) shows a clear Northwest-versus-rest contrast, plausibly related to the overall lower air pollution in the northwest region. The third FPC (12.4\%) highlights a central-U.S. pattern relative to both coasts.

\subsection{GFR Longitudinal Trajectories after Kidney Transplant}

Glomerular Filtration Rate (GFR) is a fundamental metric in nephrology, quantifying the volume of blood filtered by the glomeruli, the kidney's microscopic filtering units, per minute. This indicator is particularly important in the context of kidney transplantation. After a transplant, recipients may experience various changes in kidney function due to factors such as transplant rejection, medication effects, or the onset of chronic kidney disease. Monitoring GFR in transplant recipients thus becomes a vital component of post-operative care.
%
Our study uses a dataset compiled by \citet{liu2023functional}, who estimated GFR based on creatinine levels and associated factors.
This dataset encompasses GFR records for 143,604 recipients, tracked annually for up to seven post-transplant years. Notably, only 36.9\% of recipients have complete records for all seven years, 52.3\% have one year of data missing, and 10.7\% have five or fewer records.

The focus of our study is to perform FPCA for the GFR dataset with the post-transplant GFRs treated as irregularly observed one-dimensional functional data over time. Due to the large data size, traditional batch FPCA methods will be time-consuming. Instead, we resort to the online alternatives for FPCA including our OnlineFPCA methods and the OnlineCov method by \citet{yang2021online}.
Before analysis, we center the data by subtracting the annual mean GFR and employ a cubic B-spline basis on the interval $[1,7]$ with five evenly spaced knots to represent each individual subject. The model parameters are initialized based on the first $100$ recipients. OnlineFPCA-RAdaGrad runs with a mini-batch of size $20$ for one epoch. In dynamic tuning configured with $C=3$, $B=3$, and $W=1$, the smoothing parameter is dynamically selected from three candidates, with initialization $\{10^{-1}, 10^{-2.5}, 10^{-4}\}$, and is updated for each data block containing $200$ recipients. For OnlineCov, we also maintain three candidate bandwidths in parallel for a fair comparison.

Figure~\ref{fig:gfr-eigfun} in the supplementary materials presents the first three estimated FPCs after processing datasets of 4000, 20,000 and 100,000 recipients. As the number of data increases, the estimated FPCs given by the two methods gradually converge. Meanwhile, processing all 143,604 recipients took only around 4.2 minutes with OnlineFPCA, or 69.7 minutes when confidence intervals are computed. In comparison, OnlineCov took 565.4 minutes solely for FPC estimation. The first FPC estimated using the OnlineFPCA method remains positive throughout the seven-year period, accounting for 90.4\% of the total variation. Its corresponding score represents a weighted average of each subject's longitudinal GFR trajectory, with the first FPC serving as the weight. The second FPC is positive during the first four years and negative thereafter, explaining 7.9\% of the total variation. Its corresponding score reflects the decline in GFR after four years. The third FPC is negative between years 2 and 5 and positive at other times, contributing 1.7\% of the total variation. Its corresponding score captures the contrast in GFR between years 2-5 and the rest of the timeline.

\section{Conclusion}\label{sec:conc}

In this paper, we have proposed a comprehensive online learning framework designed to tackle the unique challenges of analyzing multidimensional functional data streams. Our methodology makes three distinct contributions to the field of functional data analysis.
First, we address the computational bottlenecks associated with multidimensional domains by formulating FPCA as a direct optimization problem on a generalized Stiefel manifold. Rather than reconstructing a high-dimensional covariance tensor, our approach effectively circumvents the curse of dimensionality that renders traditional covariance-based online methods computationally intractable for multidimensional functional data.
Second, RSGD and RAdaGrad algorithms are tailored for the manifold setting to solve this optimization problem efficiently. Combined with iterative averaging, they ensure balanced and fast convergence across leading principal components while strictly preserving geometric constraints.
Third, we integrate a dynamic tuning strategy that adapts the smoothing parameter to the streaming nature of the data.
The proposed strategy stabilizes the smoothing parameter path and is compatible with the convergence analysis of the RSGD algorithm. Under regularity conditions, we establish convergence guarantees and asymptotic normality for the averaged RSGD estimator, which further supports pointwise uncertainty quantification.

Several directions for future work can be identified. We mainly focus in this work on functional data on multidimensional Euclidean spaces. It is of practical interest to adapt the proposed approach to non-Euclidean manifolds where higher-order temporal or spatial interactions exist, for example, neuroimaging signals located over the cortical surface. Moreover, extending other batch FPCA formulations, such as those based on covariance estimation, tensor decomposition, or regression, to the online setting could enhance computational efficiency and scalability while retaining the advantages of batch methods for large-scale data. In summary, extending the online FPCA framework to more diverse models and applications holds great promise for improving real-time analysis in various domains.

\bigskip
\begin{center}
	{\large\bf SUPPLEMENTARY MATERIALS}
\end{center}

\paragraph{Supplementary document:}
The supplementary document ``supp-fpca-online.pdf'' provides background on Riemannian manifolds, computational details, proofs of theorems, and additional experiments results.

\paragraph{Code:}
A repository containing the implementation of OnlineFPCA,  code for the numerical studies, and multiple ``README.md'' files that provide instructions on how to reproduce the experiments are available at
\url{https://github.com/mynanshan/OnlineFPCA}.

\bigskip
\begin{center}
{\large\bf Disclosure Statement}
\end{center}
The authors report there are no competing interests to declare.

\setlength{\bibsep}{0pt}

\bibliographystyle{Chicago}
\bibliography{bib-fpca}


\end{document}